\documentclass[10pt,twocolumn,dutch,french,british]{iopart}
\usepackage[T1]{fontenc}
\usepackage[latin9]{inputenc}
\usepackage{babel}
\addto\extrasfrench{\providecommand{\fg}{\ifdim\lastskip>\z@\unskip\fi~\frqq}}

\usepackage{textcomp}
\usepackage{graphicx}
\usepackage[unicode=true, 
 bookmarks=true,bookmarksnumbered=false,bookmarksopen=false,
 breaklinks=false,pdfborder={0 0 1},backref=false,colorlinks=false]
 {hyperref}
\hypersetup{pdftitle={Reconnection and merging of positive streamers in air},
 pdfauthor={S. Nijdam}}
 
\makeatletter
\usepackage{iopams}
\usepackage{setstack}

\@ifundefined{showcaptionsetup}{}{%
 \PassOptionsToPackage{caption=false}{subfig}}
\usepackage{subfig}
\makeatother

\begin{document}

\title[Experiments on streamers in pure gasses and mixtures]{Probing photo-ionization: Experiments on positive streamers in pure
gasses and mixtures}

\author{S~Nijdam$^{1}$, F~M~J~H~van~de~Wetering$^{1}$, R~Blanc$^{1,2}$,
E~M~van~Veldhuizen$^{1}$ and U~Ebert$^{1,3}$}

\address{$^{1}$ Eindhoven University of Technology, Dept.\ Applied Physics\\
 P.O. Box 513, 5600 MB Eindhoven, The Netherlands}

\address{$^{2}$ \foreignlanguage{french}{Ecole Polytechnique de l'Université
d'Orléans, Orléans}, France}

\address{$^{3}$ \foreignlanguage{dutch}{Centrum Wiskunde \& Informatica},
Amsterdam, The Netherlands}

\ead{s.nijdam@tue.nl}
\begin{abstract}
Positive streamers are thought to propagate by photo-ionization of
which the parameters depend on the nitrogen:oxygen ratio. Therefore
we study streamers in nitrogen with 20\%, 0.2\% and 0.01\% oxygen
and in pure nitrogen, as well as in pure oxygen and argon. Our new
experimental set-up guarantees contamination of the pure gases to
be well below 1 ppm. Streamers in oxygen are difficult to measure
as they emit considerably less light in the sensitivity range of our
fast ICCD camera than the other gasses. Streamers in pure nitrogen
and in all nitrogen/oxygen mixtures look generally similar, but become
somewhat thinner and branch more with decreasing oxygen content. In
pure nitrogen the streamers can branch so much that they resemble
feathers. This feature is even more pronounced in pure argon, with
approximately $10^{2}$ hair tips/cm$^{3}$ in the feathers at 200
mbar; this density can be interpreted as the free electron density
creating avalanches towards the streamer stem. It is remarkable that
the streamer velocity is essentially the same for similar voltage
and pressure in all nitrogen/oxygen mixtures as well as in pure nitrogen,
while the oxygen concentration and therefore the photo-ionization
lengths vary by more than five orders of magnitude. Streamers in argon
have essentially the same velocity as well. The physical similarity
of streamers at different pressures is confirmed in all gases; the
minimal diameters are smaller than in earlier measurements.
\end{abstract}

\submitto{\JPD}

\maketitle

\section{Introduction}

\subsection{Positive streamers and photo-ionization}

Streamers are the first stage of electric break-down, when a high
voltage is applied to gas volumes~\cite{Raizer1991,Veldhuizen2000,Ebert2006a}.
The discharge can later develop into spark or lightning, but it also
can stay completely in the streamer phase. An example of a discharge
that remains essentially in the streamer phase is a sprite discharge,
a huge discharge at 40 to 90~km altitude above thunderclouds \cite{Pasko2007,Ebert2009}.
In a wide field of technical applications, the voltage pulse is intentionally
kept short to suppress inefficient gas heating during later stages
of the discharge \cite{Heesch2008,Yan2002}, and this technology builds
on streamers. Streamers enhance the electric field at their tip to
values above the breakdown value and create a region of active local
ionization dynamics; in nanosecond resolved intensified CCD-photographs,
these active areas can be seen as bright spots \cite{Blom1997,Creyghton1994,Pancheshnyi2005a}.
For further reading on streamers and sprites, we refer to a recent
cluster issue in J. Phys. D~\cite{JPDStreamer08} and to the AGU
Chapman conference on Effects of Thunderstorms and Lightning in the
Upper Atmosphere (Pennsylvania, May 2009)~\cite{Chapman09} with
its forthcoming issue in J.\ Geophys.\ Res.\ as well as to the
many original papers cited there.

While negative streamers naturally propagate through electron drift
(possibly supported by additional mechanisms), positive streamers
are typically easier to generate, but more difficult to explain~\cite{Luque2008},
as they propagate against the electron drift direction with velocities
comparable to this velocity. While Townsend~\cite{Townsend1915}
in 1915 still assumed symmetry between positive and negative charge
carriers, it soon became clear that positive ions were not suitable
for impact ionization and too slow. The commonly accepted explanation
for the propagation of positive streamers in air is photo-ionization;
this was suggested in 1935 by Flegler and Raether in Munich~\cite{Flegler1935},
by Bradley and Snoddy in Virginia~\cite{Bradley1935} and by Cravath
in California~\cite{Cravath1935}: the active ionization region emits
UV radiation that at some distance (in particular, ahead of the front)
can generate additional electron-ion-pairs. The theoretical understanding
of photo-ionization in nitrogen/oxygen mixtures was quantified by
Teich~\cite{Teich1967,Teich1967a}: The energetic electrons in the
high field region of the streamer excite certain levels of molecular
nitrogen with energies above the ionization energy of oxygen. The
levels deexcite by emission of a photon that can ionize molecular
oxygen. Teich determined the dominant wave lengths, and he also identified
quenching at higher air pressures as a mechanism suppressing photo-ionization.

There is hardly any experimental work published on measuring photo-ionization
directly in the past 40 years. Cravath~\cite{Cravath1935,Loeb1936}
already in 1935 suggested two ionization lengths of 1 and of 5~mm
in air at standard temperature and pressure, and Raether~\cite{Raether1938}
measured 5~mm in 1938 where he also investigated hydrogen and oxygen.
After the second world war, Przybylski and Teich continued this work
in Raether's lab in Hamburg and published their results in German
in 1958 and 1967~\cite{Przybylski1958,Teich1967,Teich1967a}. Penney
and Hummert~\cite{Penney1970} in Pennsylvania investigated the process
again and found in 1970 full agreement with the earlier measurements
of Przybylski and Teich, and with those of Sroka in \textquotedbl{}pure\textquotedbl{}
oxygen, \textquotedbl{}pure\textquotedbl{} nitrogen and air. The results
show that photo-ionization in air is about 1 or 2 orders of magnitude
more effective than in pure oxygen or pure nitrogen, but they still
see significant photo-ionization in both pure gasses. However, the
only information about the purity of the gasses is that \textquotedbl{}commercial-grade\textquotedbl{}
gasses have been used. This probably refers to purities in the order
of 0.1\% to 1\%. They do not discuss which mechanisms could be responsible
for the photo-ionization in any of the gasses, in contrast to Teich.
In recent years Aints \emph{et al.}~\cite{Aints2008} have used the
same method as Penney and Hummert to investigate the effects of water
content in air on photo-ionization. Their results are similar to the
results of Penney and Hummert, with small corrections for the effects
of water content.

The data of Przybylski, Teich, Penney and Hummert and the theoretical
understanding of Teich were merged by Zhelezniak \textit{et al.}~\cite{Zhelezniak1982}
into a model of photo-ionization that nowadays is used in most streamer
simulations \cite{Morrow1997,Kulikovsky2000,Naidis2006,Bourdon2007,Luque2007,Luque2008b}.

On the other hand, the reliability of the photo-ionization data and
of the resulting Zhelezniak model has been questioned; for a recent
discussion, we refer to the introduction of Nudnova and Starikovskii~\cite{Nudnova2008}.
Background ionization has been suggested as an alternative to photo-ionization;
in early simulations, photo-ionization was even replaced by background
ionization to reduce the computational complexity \cite{Dhali1987}.
The background ionization could either be due to radioactivity and
cosmic radiation, or due to left over charges at high repetition rates
of the discharge as elaborated by Pancheshnyi~\cite{Pancheshnyi2005}.
In the present experiments, high repetition rates are avoided, we
use a 1~Hz repetition frequency for all experiments.

As photo-ionization is a vital part of streamer theory, and as direct
measurements are difficult, Luque \textit{et al.}\ \cite{Luque2008b}
have suggested indirect measurements through studying the interaction
of two streamer heads propagating next to each other: they typically
would repel each other electrostatically, but could merge through
the non local effect of photo-ionization. In an attempt to confirm
this theory, Nijdam \textit{et al.} have performed experiments with
two streamers emitted from adjacent needles \cite{Nijdam2009}, but
a common parameter regime has not yet been explored for theory and
experiment, and the photo-ionization could not yet be investigated
along these lines. Kashiwagi and Itoh have shown that UV and VUV radiation
from a surface streamer discharge can trigger a synchronous streamer
discharge~\cite{Kashiwagi2006}. They have found that radiation around
115~nm is most effective, which is somewhat higher than the limit
of 102.5~nm that is associated with nitrogen-oxygen photo-ionization.
However, in the case of surface discharges, photo-electron emission
from the insulator surface itself can replace photo-ionization of
oxygen molecules in bulk streamer discharges. The photo-electron emission
from the surface can occur at higher wavelengths than the bulk photo-ionization
of oxygen molecules.

Here we follow a different experimental track. According to Teich's
photo-ionization mechanism in nitrogen-oxygen mixtures, the density
of emitted photons is proportional to the nitrogen concentration,
and the absorption lengths of the photons are inversely proportional
to the oxygen concentration. If positive streamers in air indeed propagate
through this mechanism, one would expect that their properties change
when the ratio of nitrogen and oxygen is changed. The main purpose
of this paper is therefore to investigate positive streamers in varying
nitrogen:oxygen mixtures experimentally. To set the limits, we also
investigate streamers in pure nitrogen and in pure oxygen, and for
comparison, we also investigate them in pure argon.

\subsection{Previous experiments on streamers in different gases}

\subsubsection*{Streamers in nitrogen-oxygen mixtures}

Streamers in varying nitrogen-oxygen-mixtures have been investigated
before by Yi and Williams~\cite{Yi2002}, Ono and Oda~\cite{Ono2003},
and Briels \textit{et al.}~\cite{Briels2008a,Briels2008b}.

Yi and Williams~\cite{Yi2002} use a 130~mm plane-plane with protruding
point geometry, where the protruding point is a 3.2~mm diameter rod
ground to a sharp tip (radius $\approx$100~\textmu{}m). They use
both positive and negative voltages pulses with amplitudes between
70 and 130~kV and rise times of order 50~ns. Nitrogen-oxygen mixtures
with oxygen fractions between \textquotedbl{}0\%\textquotedbl{} and
15\% are investigated. The purity of their pure nitrogen is not specified
but it is presumably significantly below 0.1\%, as this is the oxygen
content of their next gas mixture. They claim that their measurements
strongly suggest that photo-ionization plays an important role because
(especially positive) streamers propagate faster at higher oxygen
concentrations. For high voltages (>100~kV), the propagation velocity
of positive streamers increases with roughly a factor five when going
from their pure nitrogen to 10\% oxygen in nitrogen. In negative streamers
the same concentration change leads to a velocity increase of less
than 40\%.

Ono and Oda~\cite{Ono2003} also use a point-plane geometry, but
with a 13~mm gap. Their tips are made by cutting a 0.3~mm stainless
steel wire and have no well defined tip profile. They apply positive
voltage pulses with amplitudes between 13 and 37~kV, rise times in
the order of tens of nanoseconds and durations of a few hundred nanoseconds
on nitrogen-oxygen mixtures containing oxygen fractions between \textquotedbl{}0\%\textquotedbl{}
and 20\%. Again, the purity of the nitrogen used in the experiments
is not specified. Their next purest mixture contains 0.2\% oxygen.
They claim that propagation velocity, diameter and shape of the streamers
are strongly influenced by the oxygen concentration. No measured value
varied by more than a factor of five when changing the oxygen concentration
from 20\% to 0\%. The streamer diameter increases from 0.2-0.4~mm
in pure nitrogen to more than 1~mm in air. At 18~kV, the propagation
velocity increases from $\sim2\cdot10^{5}\,\mathrm{m/s}$ in pure
nitrogen to $5\cdot10^{5}\,\mathrm{m/s}$ in air.

Briels \textit{et al.}~\cite{Briels2008a,Briels2008b} have measured
in air, a mixture with 0.2\% oxygen in nitrogen and \textquotedbl{}pure
nitrogen\textquotedbl{}. They use 5 to 30~kV positive pulses with
rise times of order 20 -- 200~ns in point-plane gaps of 40 and 160~mm
at pressures between 100~mbar and 1~bar. They have shown that at
lower oxygen concentrations, streamers branch and zigzag more, they
are brighter and thinner (about 40\% thinner in pure nitrogen than
in air). They have not found a clear effect of oxygen concentration
on propagation velocity; at 1 bar they are lowest in pure nitrogen,
while at lower pressures they are the same or highest in pure nitrogen.
However, the set-up that was used in these experiments is not designed
for high purity gas handling. Therefore, it is not known what the
exact purity of the pure nitrogen was in all of the experiments described
by Briels \textit{et al.}

In order to better investigate the effects of low oxygen concentrations,
here we will present results in which the purity of the gasses can
be guaranteed to ppm (parts per million) levels.

\subsubsection*{Streamers in argon}

Aleksandrov \emph{et al.} \cite{Aleksandrov2001} have performed experiments
on streamers in pure argon (99.99\% purity) and a mixture of 1 to
5\% oxygen in argon, all at atmospheric pressure. They present streak
photography of discharges in a 250~mm rod-plane gap with 10 to 60~kV
voltage pulse with 1~\textmu{}s rise time. They conclude that a non-thermal
mechanism of streamer breakdown in the case of argon discharges gives
way to leader breakdown when 1\% or more of oxygen is added. This
addition also leads to a decay of the streamer channel that is an
order of magnitude faster and therefore a noticeable higher electric
field is needed for the streamers to bridge the gap. Their simulations
show that this can be explained by the quenching of excited argon
molecules by oxygen molecules. Van Veldhuizen and Rutgers~\cite{Veldhuizen2002}
have used air and argon. They found that streamers in argon show less
branching and can only be produced in a narrow voltage range.

Measurements similar to the ones presented in this paper on two planetary
gas mixtures (Venus and Jupiter) are presented in a paper by Dubrovin
\emph{et al.} \cite{Dubrovin2009}.

\subsection{Advantages of short pulses}

Past measurements by Briels \emph{et al.~}\cite{Briels2006,Briels2008,Briels2008a,Briels2008b}
and others~\cite{Aleksandrov2001,Yi2002,Ono2003} have been performed
with voltage pulses with durations of many hundreds of nanoseconds
up to many microseconds and with risetimes of tens of nanoseconds
or more. This approach has two disadvantages: firstly, the slow risetime
means that the streamers may initiate at a significantly lower voltage
than the reported maximum voltage. Therefore, they may not exhibit
the properties (like diameter and propagating velocity) that are representative
for this voltage, but for a lower voltage. Secondly, a long pulse
duration may lead to a streamer-to-spark transition. Because a spark
is usually much brighter than a streamer, it may damage the photo-cathode
of an ICCD camera. 

One way to overcome these two problems is to use a power supply based
on a Blumlein pulser~\cite{Smith2002,Liu2006}. With such a circuit,
it is possible to make pulses with a well defined duration (depending
on the cable length) and a fast risetime (order 1 to 10~ns). Briels
\emph{et al.} have investigated the effect of risetime on streamer
properties. We extend this to lower risetimes.

\subsection{Content of the paper}

In this paper we present experimental measurements on streamer discharges
in different gas mixtures, with an emphasis on nitrogen/oxygen mixtures,
focusing on photo-ionization. We will repeat some measurements by
Briels \emph{et al.} \cite{Briels2008b}, but with a better defined
purity of the gasses and better optics to measure streamer diameters.
We also extend the measured range to nitrogen of 1~ppm purity and
to argon and oxygen of 10~ppm purity and we discuss implications
for the propagation mechanism. We measure at various pressures to
further test streamer similarity laws.

\section{Experimental set-up.}

\begin{figure}
\subfloat[]{\includegraphics[clip,height=5cm]{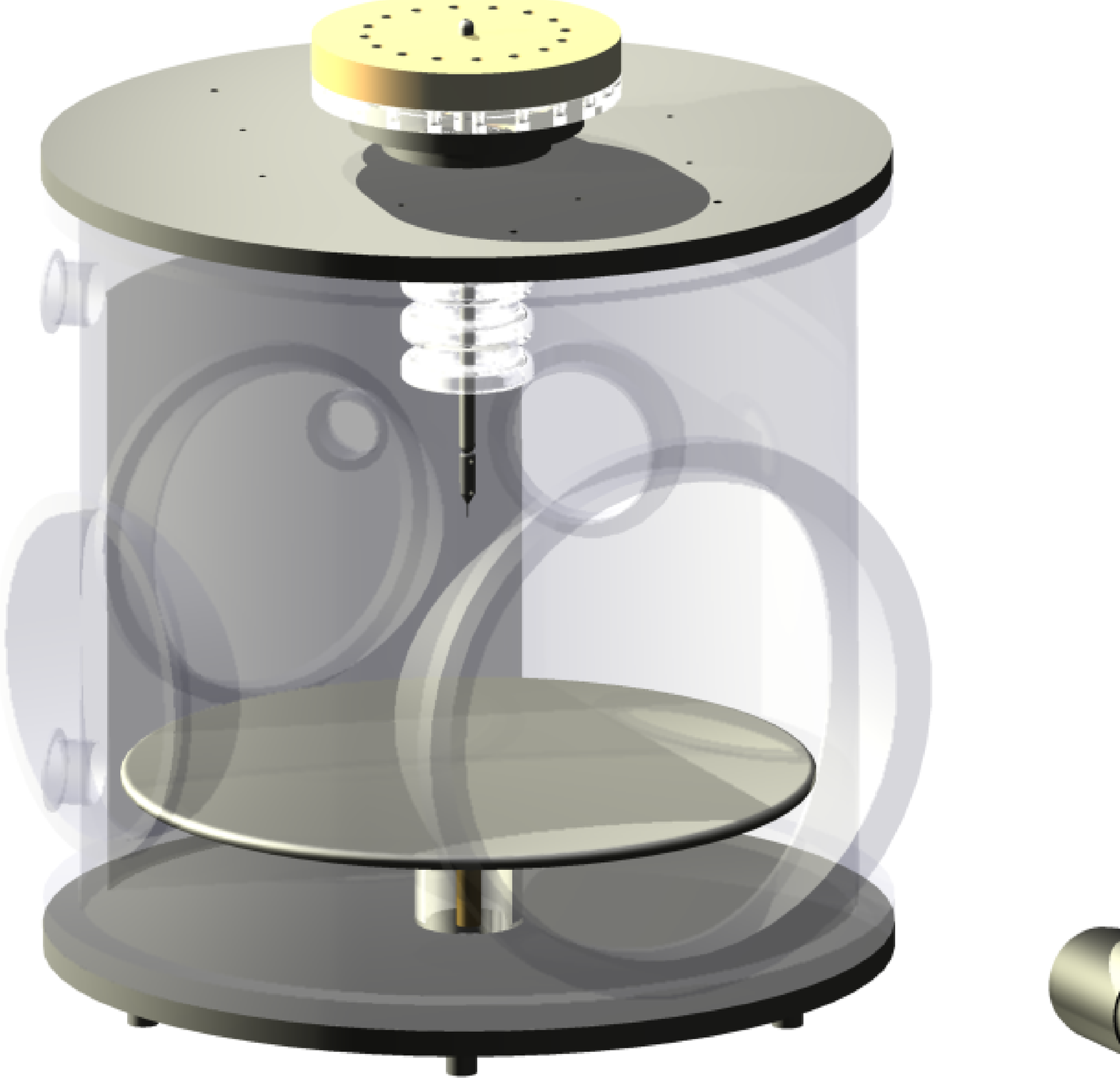}}\subfloat[]{\includegraphics[clip,height=5cm]{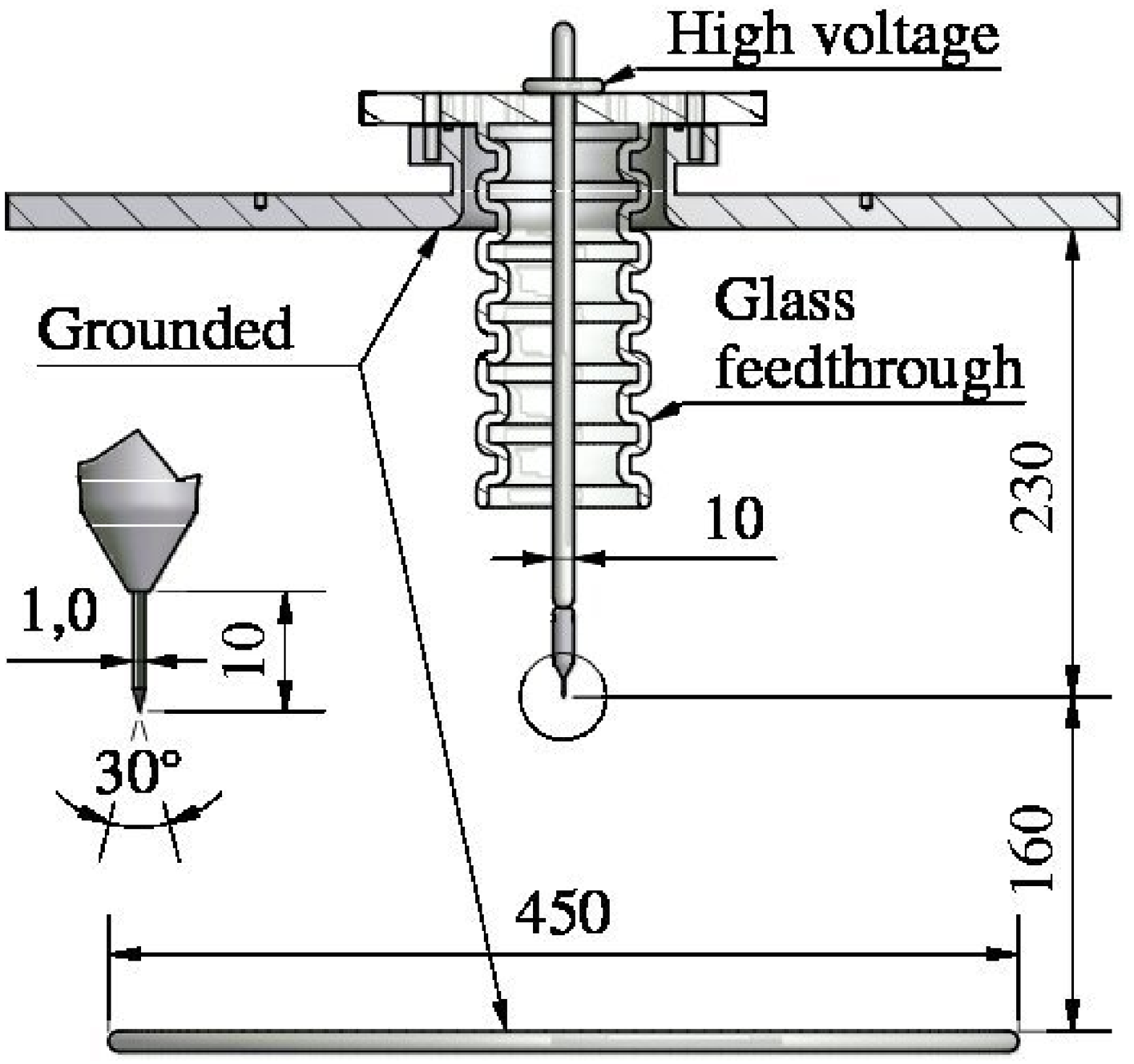}

}

\caption{\label{fig:Overview-setup}Overview (a) and schematic view (b) of
the new high purity vacuum vessel with the ICCD camera. The wall of
the vessel has been rendered transparent in the overview so that the
anode tip and cathode plane are clearly visible. In the schematic
view, dimensions are given in mm. The bottom of the tip holder and
the tungsten anode tip are enlarged on the left side of the image.}

\end{figure}

We have built a set-up that is specifically designed to ensure the
purity of the enclosed gasses. For this reason, the set-up can be
baked to reduce out-gassing, it contains no plastic parts, except
for the o-ring seals and it stays closed all of the time. When not
in use, the set-up is pumped down to a pressure of about $2\cdot10^{-7}$~mbar.
The leak rate (including out-gassing) under vacuum is about $8\cdot10^{-7}$~mbar~min$^{-1}$,
which is considerably lower than the $5\cdot10^{-3}$~mbar~min$^{-1}$
reported by Yi and Williams~\cite{Yi2002}. We have used a helium
leak tester to check for any leaks, but have not found any.

During use, the gas inside the set-up is flushed with such a flow
rate that all gas is replaced every 25~minutes. The absolute flow
rate is controlled by a mass-flow controller and depends on pressure.
The flow rate is sufficiently high that the contamination caused by
the leak/out-gassing rate is significantly below 1~ppm for all pressures
used. We use mixtures of O$_{2}$ and N$_{2}$$ $ that are pre-mixed
by the supplier. According to the specifications, the amount of contamination
is below 1~ppm. The mixtures we have used contain nitrogen with 20\%,
0.2\%, 0.01\% and <0.0001\% oxygen (volumetric fractions). We refer
to the 20\%~oxygen mixture as artificial air. Besides these mixtures,
we have also used pure oxygen and pure argon, both with not more then
10~ppm impurity.

The vacuum vessel contains a sharp tungsten tip, placed 160~mm above
a grounded plane. A schematic drawing of the vacuum vessel with the
camera is given in figure~\ref{fig:Overview-setup}. During a measurement,
a voltage pulse is applied to the anode tip. The vacuum vessel is
located inside a Faraday cage in order to protect the camera and other
sensitive electronics from the electromagnetic pulses of the pulse
source.\\

A propagating streamer head emits light and the path of these heads
can therefore be imaged onto a camera with a lens. Because the emitted
light is often weak an intensified CCD (ICCD) is needed. The intensifier
also enables us to take images with very short (nanosecond) exposure
times. These images only show a small section of the streamer propagation
and can therefore be used to measure the velocity of the streamers
\cite{Briels2008}.

In our case, streamer discharges are imaged by a Stanford Computer
Optics 4QuikE ICCD camera. We use two different lens assemblies: a
Nikkor UV 105~mm f/4.5 camera lens mounted directly on the camera
and a 250~mm focal length, 50~mm diameter achromatic doublet on
an optical rail. The latter is used to zoom in on a specific region
of the discharge in order to measure small streamer diameters. We
use a window of conducting ITO (Indium tin oxide) glass as part of
the Faraday cage to protect the camera against the high voltage pulses.

In the images presented here, the original brightness is indicated
by the multiplication factor \textit{Mf}, similar to what Ono and
Oda introduced in~\cite{Ono2003}. This value is a measure for the
gain of the complete system, it includes lens aperture, ICCD gain
voltage and maximum pixel count used in the false-colour images. An
image with a high \textit{Mf} value, is in reality much dimmer than
an image with similar colouring, but with a lower \textit{Mf} value
(the real brightness is proportional to \textit{Mf}$^{-1}$ for an
image with the same colours). We have normalized the \textit{Mf} value
in such a way that the brightest image presented here has an \textit{Mf}
value of 1.

\subsection{Pulse sources}

\begin{figure}
\includegraphics[width=1\columnwidth]{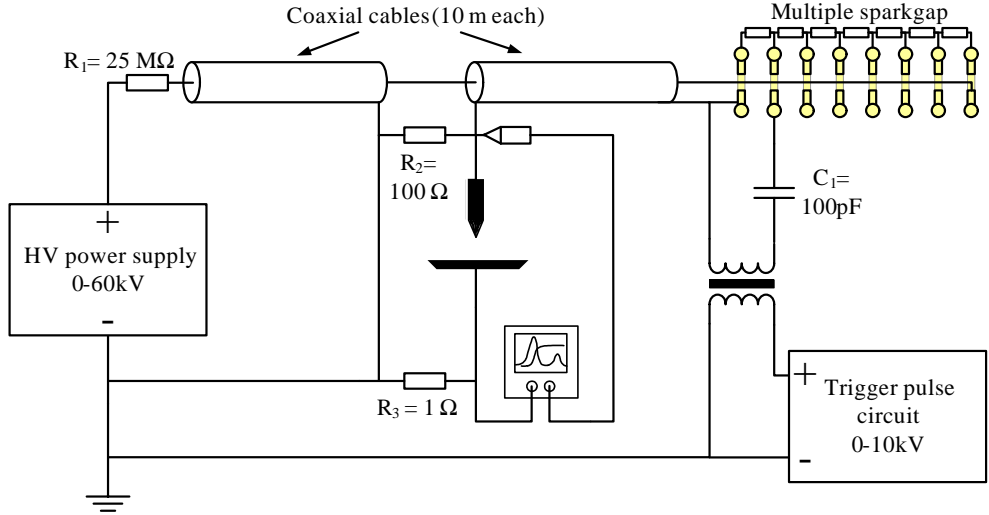}\caption{\label{fig:Blumleinpulser}Schematic circuit of the Blumlein pulser
design. The trigger pulse circuit has been omitted for sake of simplicity.}

\end{figure}

We have used two different electric circuits to produce a voltage
pulse: the so-called C-supply and a newly designed circuit based on
a Blumlein pulser with a multiple sparkgap.

The C-supply consists of a 1~nF capacitor that is charged by a high
voltage negative DC source. Now a trigger circuit triggers a spark
gap, which acts as a fast switch. The capacitor is discharged and
puts a positive voltage pulse on the anode tip. This pulse has a 10\%
-- 90\% rise-time of at least 15~ns and a fall time of about 10~\textmu{}s,
depending on discharge impedance and choice of resistors. This circuit
is treated extensively in \cite{Briels2006}.

\begin{figure}
\includegraphics[width=1\columnwidth]{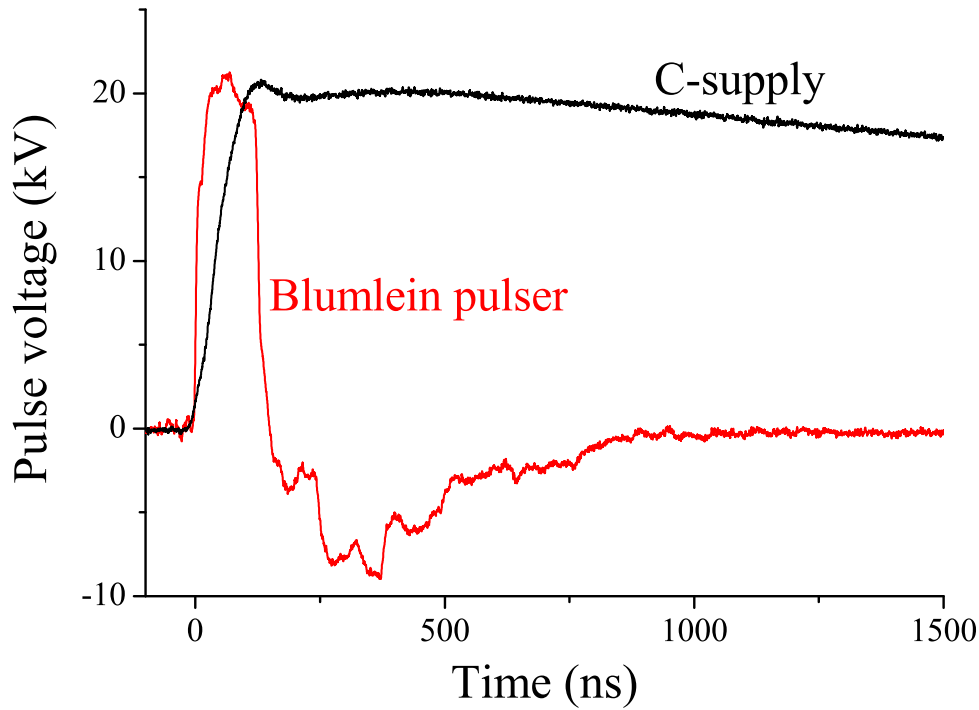}\caption{\label{fig:Comparison-pulse-shapes}Comparison of the voltage pulses
produced by the two different pulse sources. Note that the duration
of the pulse from the C-supply depends largely on the impedance of
the discharge and can therefore vary significantly according to gas
type and pressure. Because the Blumlein pulser contains a low-ohmic
resistor in parallel to the discharge, it is hardly influenced by
the discharge impedance.}

\end{figure}

\label{Blumlein-pulser}The Blumlein pulser is a new design based
on existing knowledge about such circuits~\cite{Smith2002,Liu2006}.
It produces a more or less rectangular pulse with a short rise time
and a fixed duration. The present design allows for a rise time of
about 10~ns and a pulse duration of 130~ns, see figure~\ref{fig:Comparison-pulse-shapes}.
This is achieved by charging two coaxial cables (see figure~\ref{fig:Blumleinpulser})
with a DC voltage source. When the multiple sparkgap switch \cite{Liu2006a}
at the end of one of these cables is closed, a voltage pulse will
travel along the cable mantles. This voltage pulse will arrive at
the anode tip. After traversing the second cable back and forth, the
voltage pulse at the anode tip will be nullified. We see only one
reflected pulse occurring after the first pulse (starting at $t\approx250$~ns).
This indicates that our system is quite well matched.

This new design has three advantages compared to the C-supply: The
faster rise time ensures that the voltage during initiation and early
propagation is closer to the reported maximum voltage; therefore thicker
streamers are generated~\cite{Briels2006}. The short duration of
the pulse enables us to use overvolted gaps without the risk of spark
formation (this can be dangerous for the camera). The short duration
also helps to ensure that only primary streamers are produced. This
makes it much easier to perform spectroscopic measurements on only
these primary streamers; we can use long integration times without
capturing other phenomena like secondary streamers or glow discharges.
However, this last advantage is only valid if the streamers do not
bridge the gap long before the end of the pulse duration. In our measurements
this is not always the case, as is shown in figure~\ref{fig:Movie25mbar}.
The length of the pulse can be varied by using other lengths for the
two coaxial cables, although this takes some time and is not easy
to implement during a measurement series.

\subsection{Measuring streamer diameter and velocity}

From the images taken, the streamer diameter is determined with the
following method:
\begin{itemize}
\item A straight streamer channel section is selected.
\item In this section, several perpendicular cross sections of the streamer
are taken.
\item These cross sections are averaged so that they form one single cross
section.
\item The streamer diameter is determined as the full width at half maximum
(FWHM) of the peak (streamer intensity).
\end{itemize}
A diameter of at least 10 pixels in the image is required to be sure
that the streamer diameter is measured correctly and camera artifacts
are negligible. For higher pressures (with thinner streamers), we
have used the doublet lens in order to zoom in on a small specific
region of the discharge.\\

\begin{figure}
\includegraphics[clip,width=9cm]{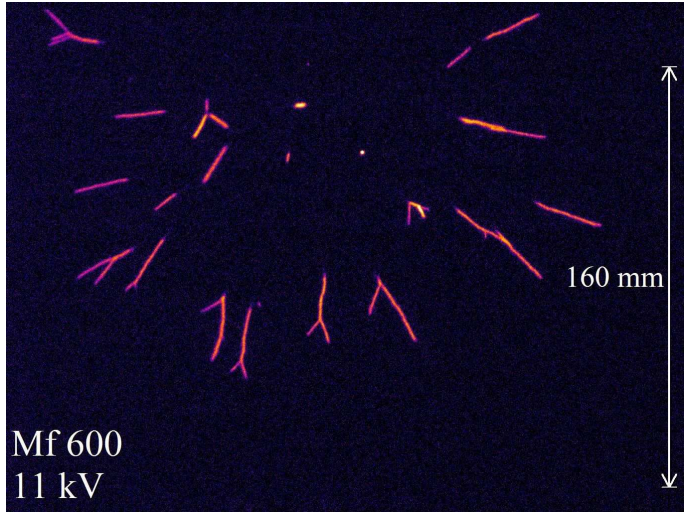}\caption{\label{fig:Velocity-example}Example of an ICCD image used to determine
streamer propagation velocities in nitrogen-oxygen mixtures. The velocity
is calculated by dividing the length of the longer streamer sections
by the exposure time. This example shows an image of the 0.2\% oxygen
in nitrogen mixture at 200~mbar with an exposure time of 300~ns.}

\end{figure}
The streamer propagation velocity has been measured with two different
methods. The first method is as follows: we take short exposure images
of streamers while they propagate in the middle of the gap (see figure~\ref{fig:Velocity-example}).
We then choose the thinner, straighter and longer streamer sections
in each picture. The thinner streamers are insured to be in focus,
the longer streamers are assumed to propagate almost in the photograph
plane. The length of each such streamer section is measured (we use
a similar FWHM method as for the diameter measurement), and the velocity
is calculated as the ratio between length and exposure time, corrected
for head size, depending on exposure time. Streamer sections that
contain a branching event are ignored. 

This method could be improved by using stereo-photographic techniques
as demonstrated in \cite{Nijdam2008,Nijdam2009}. However, for sake
of simplicity, we have chosen not to do this in the measurements presented
here.

\label{Velocity-technique}In the case of measurements on streamer
velocity in argon, this first method does not work because of the
long lifetimes of the excited states of argon. Even with a very short
integration time and a delay long enough to start observing when the
streamers are halfway through the gap, one will still see the entire
trail between the streamer head and the anode tip. Therefore we have
to measure the velocity by measuring the distance travelled by the
fastest streamer heads as function of time. When we now differentiate
this curve we have the velocity as function of time. For this method
to work we need to have little jitter in the streamer initiation.
This method has previously been used by Winands \emph{et al.} \cite{Winands2008a,Briels2008}
to measure streamer propagation velocities.

With both methods one should keep in mind that streamer velocity is
not constant in the strongly non uniform field of our point-plane
discharge gap (see e.g. figure 7 in~\cite{Briels2008b} and accompanying
discussion there). However, after roughly half of the gap, the velocity
does not change much, except when the streamers get close (a few streamer
diameters) to the cathode plate. Therefore, we have chosen to use
images in which the streamers are roughly halfway into the gap (for
both methods). We have verified that both methods described above
give the same results for gasses that support the first method.

More information about the discharge vessel, experiment timing, imaging
system and measurement techniques can be found in \cite{Briels2008,Briels2008b,Nijdam2009}.

\section{Results and discussion}

\subsection{Images and general morphology}

\begin{figure*}
\includegraphics[height=0.7\paperheight]{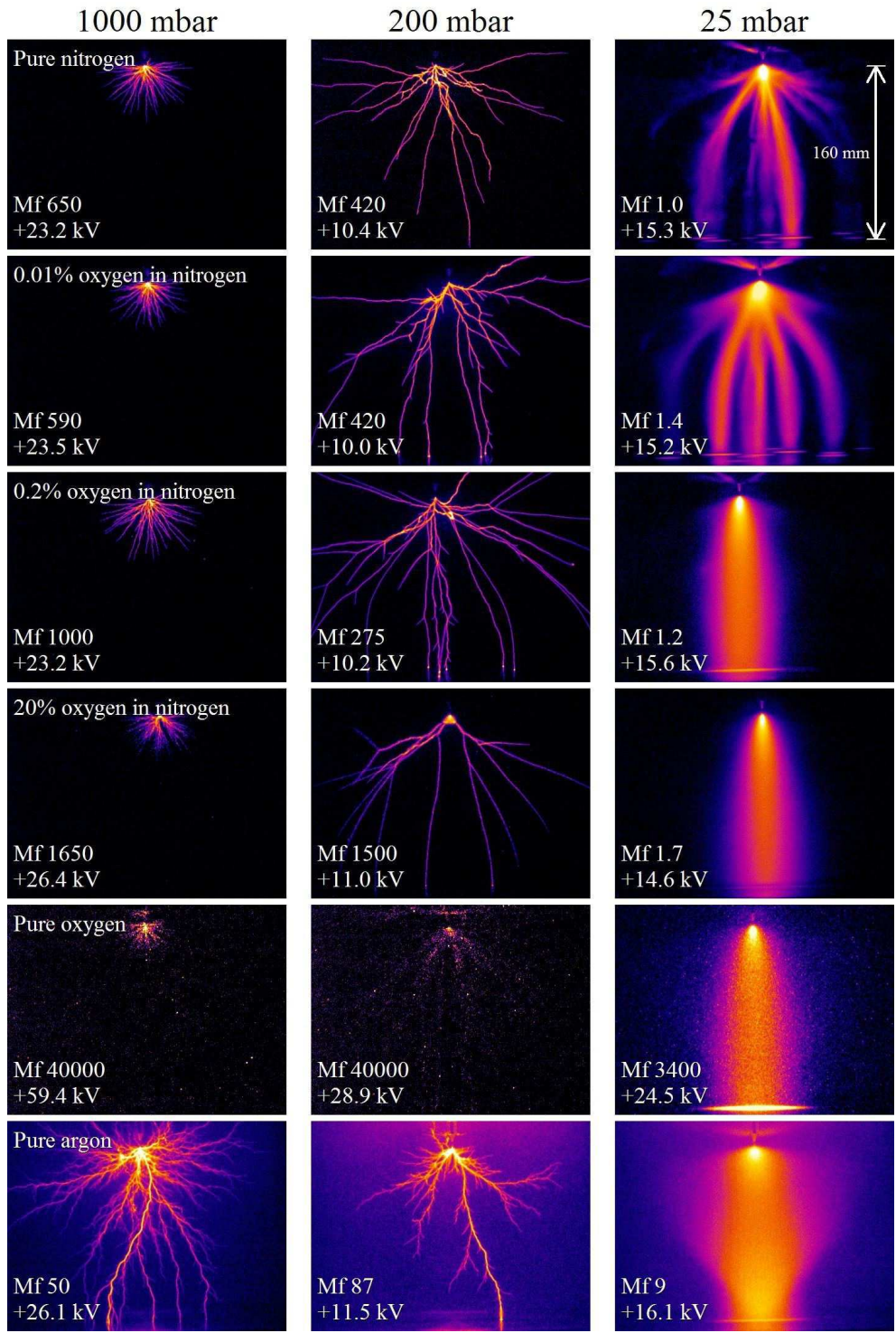}

\caption{\label{fig:Overview-images}Overview of streamer discharges produced
with the C-supply for all six gas mixtures used (rows), at 1000, 200
and 25~mbar (columns). All measurements have a long exposure time
and therefore show one complete discharge event, including transition
to glow for 25~mbar. The multiplication factor (\textit{Mf}) gives
an indication of the real intensity of the discharge. The white arrow
at the top right image indicates the vertical position of the anode
point and the cathode plane. Note that the voltages used are significantly
higher for the oxygen measurements. At 25~mbar the discharge is partly
in the glow-phase instead of the streamer-phase. Images at other voltages
can be found in the multimedia attachments.}

\end{figure*}

We investigate the general morphology of the streamers by means of
ICCD camera images. Previous time resolved photography in a similar
set-up has established the following sequence of discharge evolution:
first a glowing ball or initiation cloud appears at the needle electrode,
the ball extends and transits into an expanding glowing shell that
eventually becomes unstable and breaks up into many simultaneously
propagating streamers \cite{Briels2008b,Briels2008c}. The sizes of
the initiation cloud and glowing shell scale inversely with pressure.
At high pressures (1000~mbar in our case), they are so small that
they can not be distinguished in our images. At low pressures, or
when the size of the gap is very small, the imitation cloud and glowing
shell may extend over more than half the gap. Pressure, gas mixture,
gap length, voltage and voltage rise time determine whether a single,
few or many streamers emerge from the glowing shell.

In gas mixtures containing nitrogen, most radiation is produced by
molecular nitrogen. It is well known that the streamer heads only
emit light for a very short time (less than 2~ns) in these mixtures
\cite{Blom1994,Shcherbakov2007,Briels2008}. We can assume that the
atoms or molecules that emit the radiation have not moved significantly
between excitation and emission. For example, the thermal velocity
of a nitrogen molecule at room temperature is about 500~m/s. This
means that the maximum distance it can travel within 10~ns is 5~\textmu{}m
if it is not scattered on its path; This is clearly below our resolution.
Therefore, the image represents a mapping of the production of the
relevant excited states in nitrogen. Some time resolved pictures will
be shown later in this paper. First we present time integrated ICCD
pictures of such single discharge events\\

As was discussed already above, in argon the streamer channels remain
bright tens to hundreds of nanoseconds after the streamer head has
passed. However, even in this case the maximum distance travelled
by an excited atom or molecule before it decays will still be smaller
than the pixel size of our camera. Therefore, in all cases, the images
represent a map of the locations where the excited molecules or atoms
are created and the images are not influenced by diffusion of these
excited species after excitation.

\subsubsection*{Nitrogen-oxygen mixtures and pure nitrogen}

Figure~\ref{fig:Overview-images} shows an overview of streamers
made with the C-supply in the six different gas mixtures. The general
morphology of the N$_{2}$:O$_{2}$ mixtures and pure nitrogen is
very similar, although it is clear that, especially at low pressures,
there are more branches in pure nitrogen and the 0.01\% mixture than
in the other two mixtures. This has also been observed by Ono and
Oda~\cite{Ono2003}. Their streamer channels become marginally straighter
for higher concentrations of oxygen.

One striking feature in the N$_{2}$:O$_{2}$ mixtures and the pure
nitrogen is the maximal length of the streamers at 1000~mbar; the
streamers are longest for the 0.2\%~O$_{2}$ mixture. However, the
interpretation is not straightforward. In these images the exposure
time of our camera was about 2~\textmu{}s, and the decay time ($1/e$)
of the voltage pulse was about 6~\textmu{}s. On the other hand, Briels
\emph{et~al.}~\cite{Briels2008b} in their \textquotedbl{}pure nitrogen\textquotedbl{}
have found much longer streamers under similar conditions that propagated
for more than $4\,\mathrm{\mu s}$ after the start of the pulse. In
those measurements the decay time of the voltage pulse was probably
much longer than $6\,\mathrm{\mu s}$ because they used a higher parallel
impedance in the C-supply circuit (the exact value is unclear). The
lengths of our streamers in pure nitrogen therefore can be determined
either by the exposure time of the camera or by the decay time of
the voltage pulse. In air, streamers stay short both in Briels' and
in the present measurements. This is probably due to the conductivity
loss inside the streamer channel due to the attachment of electrons
to oxygen molecules. In air, this loss mechanism is much stronger
than in pure nitrogen with a small amount of oxygen contamination.

The images of the 0.2\%~O$_{2}$ and 0.01\%O$_{2}$ are the brightest
at 1000~mbar and (less so) at 200~mbar. This confirms results of
Ono and Oda~\cite{Ono2003}. One should be careful when comparing
exact \textit{Mf} values of our measurements. We have not conducted
a statistical study into the brightness of the streamer channels.

\subsubsection*{Pure oxygen}

The images of pure oxygen are clearly more noisy than all others because
of the low light intensity of the streamers, which leads to high multiplication
factors. In order to make the images we had to increase the gain of
the ICCD camera to close to its maximum level. The intensity of the
streamers in pure oxygen is 2 to 3 orders lower than for the streamers
in other gasses. When the voltage of the pure oxygen measurement would
have been around 25~kV as for the other gasses, this difference would
be even larger. However, at 25~kV nothing could be seen and therefore
we have used a 59.4~kV image. At 200~mbar, the situation was similar.

Note that the measured intensity also depends on the emission spectrum,
as a camera always has a wavelength dependent response. Our camera
is sensitive between 200 and 800~nm. However, because of the ITO
glass in our Faraday cage, radiation below 300~nm is cut off. 

The low intensity of the streamers in oxygen can be explained by the
lack of emission lines from oxygen in the sensitive region of the
camera. The strongest line at 777~nm is close to the limit of the
camera sensitivity curve. This makes it nearly impossible to do any
quantitative research on streamers in pure oxygen with our camera.
The images at pressures above 50~mbar are not suitable for diameter
or velocity measurements. At high pressures it is often even difficult
to determine if streamers are present or not. Therefore we will not
present any other data on pure oxygen streamers.

\subsubsection*{Pure argon}

Pure argon has a significant light emission intensity and yields good
quality camera images. In comparison to the other gasses, it branches
much more, but most branches are very short. It also develops easily
into a spark, which means that we have to take care that the voltage
pulse remains short enough to avoid this (we did not use the Blumlein
pulser with argon). As was discussed already above, in argon the streamer
channels remain bright tens to hundreds of nanoseconds after the streamer
head has passed. However, even in this case the maximum distance travelled
by an excited atom or molecule before it decays will still be smaller
than the pixel size of our camera. Therefore, in all cases, the images
represent a map of the locations where the excited molecules or atoms
are created and the images are not influenced by diffusion of these
excited species after excitation. In the 1000 and 200~mbar images
we can already see that one channel is much brighter than the others.
The relatively bright channel is the reason for the relatively low
\textit{Mf} values for argon at these pressures. This near-sparking
behaviour is frequently observed in the argon discharges.

The easy sparking and heavy branching is probably due to the fact
that argon is a noble gas without rotational or vibrational degrees
of freedom, as we will discuss in more detail in section~\ref{Feather-interpretation}.
Therefore there are few inelastic collisions at low electron energies,
and hence impact ionization becomes effective at much lower field
strengths $E$ than in nitrogen or oxygen. The impact ionization coefficient
$\alpha(E)$ is more than 2 orders of magnitude larger for argon than
for nitrogen at fields below 30~kV/cm at atmospheric pressure \cite{Siglo}.\\

\begin{figure*}
\includegraphics[clip,width=1\columnwidth]{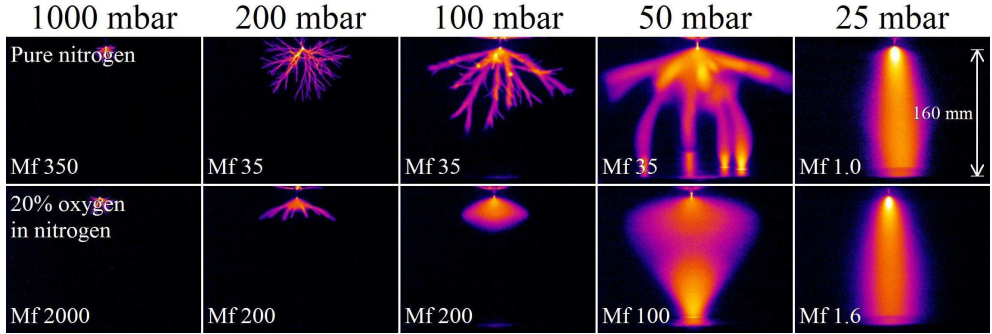}

\caption{\label{fig:Blumlein-images}Overview of streamer discharges produced
with a 20~kV pulse by the Blumlein pulser ($\approx$130~ns pulse
length) for pure nitrogen (top row) and artificial air (bottom row),
at five different pressures. All measurements have a long exposure
time and therefore show the whole discharge.}

\end{figure*}

\subsubsection*{Results from the Blumlein pulser}

The differences between artificial air and pure nitrogen can be studied
better with the Blumlein pulser than with the C-supply. We recall
that the Blumlein pulser has a faster rise time and a shorter pulse
duration than the C-supply. This allows us to use the same voltage
at all pressures without danger of provoking a spark. Hence in figure~\ref{fig:Blumlein-images},
the electric fields $E$ are the same up to modifications due to the
presence of the discharge while the reduced fields $E/N$ increase
with decreasing density $N$.

This figure shows that at 25~mbar and 1000~mbar the images for pure
nitrogen and artificial air are quite similar, but all intermediate
pressures show clear differences between the two gasses. The figure
shows that at 25~mbar and 1000~mbar the images for pure nitrogen
and artificial air are quite similar, but all intermediate pressures
show clear differences between the two gasses. Again, artificial air
has fewer and thicker streamers, as well as a much larger initiation
cloud or glowing shell (see discussion at the beginning of this section).
Because of the limited duration of the pulse, no streamers emerge
from the glowing shell in artificial air at 100~mbar. This limited
duration is also the reason for the short streamer lengths at pressures
above 50~mbar for both gasses. We have used a voltage of 20~kV in
all measurements in figure~\ref{fig:Blumlein-images}; this voltage
is higher than in figure~\ref{fig:Overview-images} for all pressures
below 1000 mbar.\emph{}\\

The differences in appearance between air and pure nitrogen with the
Blumlein pulser are also more pronounced than found by Briels \emph{et
al.}~\cite{Briels2008b}. This can again be attributed to the effect
of a higher voltage and faster rise time in our case, but also to
the higher purity that can be achieved in the new set-up. It seems
that the thinnest or minimal streamers are similar in the different
mixtures, but that air produces thicker streamers more easily.

The larger \textquotedbl{}initiation cloud\textquotedbl{} in artificial
air can also be observed in the 200~mbar images from figure~\ref{fig:Overview-images},
when comparing this with the lower-oxygen concentration mixtures.

In figure~\ref{fig:Blumlein-images}, in pure nitrogen at 25~mbar
there is only one channel instead of the multiple channels in figure~\ref{fig:Overview-images}.
This is caused by the higher voltage (20~kV versus 15.3~kV) and
faster rise time ($\sim$10~ns versus $\sim$50~ns) from the Blumlein
pulser compared to the C-supply. Such a high voltage would lead to
a spark when using the C-supply.

The shapes of all discharges at pressures of 50~mbar and lower correspond
to the shapes found by Rep'ev and Repin~\cite{Rep'ev2006} in discharges
in a 100~mm point plane gap in air at atmospheric pressure with 220~kV
pulses. So their reduced background field ($E/N$) is quite similar
to ours. They use a bullet-shaped electrode with a tip radius of 0.2~mm.

\subsubsection*{Initiation delay and jitter}

We have not observed any effects of gas mixture on initiation delay
or jitter as reported by Yi and Williams~\cite{Yi2002}. In all cases,
the streamer initiation jitter was limited to less than 10~ns. This
can probably be attributed to our electrode holder geometry, which
induces more field enhancement at the tip, our fast voltage rise times
and our overvolted gaps. The protrusion-plane electrode of Yi and
Williams has a tip radius of about 100~\textmu{}m and protrudes only
10~mm from a plane. Our tip has a radius of about 15~\textmu{}m
and is mounted in an elongated holder (see figure~\ref{fig:Overview-setup}).
Van Veldhuizen and Rutgers~\cite{Veldhuizen2002} have shown that
such a difference in geometry of the holder and tip can have a large
influence on streamer initiation and propagation. Therefore, initiation
was always fast in our case and we did not observe differences between
the gas mixtures.

\subsection{Feather-like structures}

\begin{figure}
\includegraphics[width=8cm]{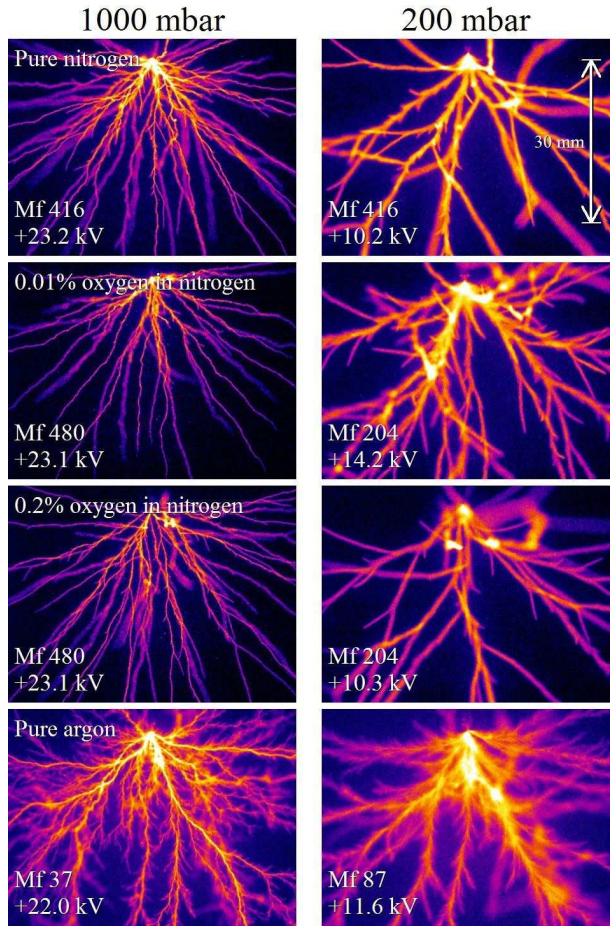}

\caption{\label{fig:Zoomed-images}Zoomed images produced with the C-supply
for four different gasses (rows), at 1000 and 200 mbar (columns) around
the anode tip region (see length indication at top right image). All
other settings are similar to figure~\ref{fig:Overview-images}.
Note that the image for 0.01\%~oxygen at 200~mbar has a deviating
voltage.}

\end{figure}
Figure \ref{fig:Zoomed-images} shows images from similar discharges
as in figure \ref{fig:Overview-images} but now zoomed to the region
closest to the anode tip. The zooming is achieved by moving the camera
closer to the experiment. We omitted images from pure oxygen and from
artificial air. The former is omitted because it is hardly visible,
for the latter we unfortunately do not have any zoomed measurement
images available at 200~mbar and voltages between 10 and 15~kV.
We do have zoomed images of artificial air at 1000~mbar and 23~kV.
These are very similar to the corresponding images of the 0.2\% oxygen
mixture. For 200~mbar however, we do not expect that is the same,
as the unzoomed images from figure~\ref{fig:Overview-images} already
show large differences between these two gas mixtures, also close
to the electrode tip region.

We observe again that for lower oxygen concentrations, the streamers
branch more often. In pure nitrogen at 200~mbar they form shapes
resembling feathers. In argon, these feather-like structures are even
more pronounced, especially at 200~mbar. The fact that there are
no straight, smooth streamer channels in the argon discharges makes
it impossible to determine unambiguous streamer widths in this gas.\\

\begin{figure}
\includegraphics[width=8cm]{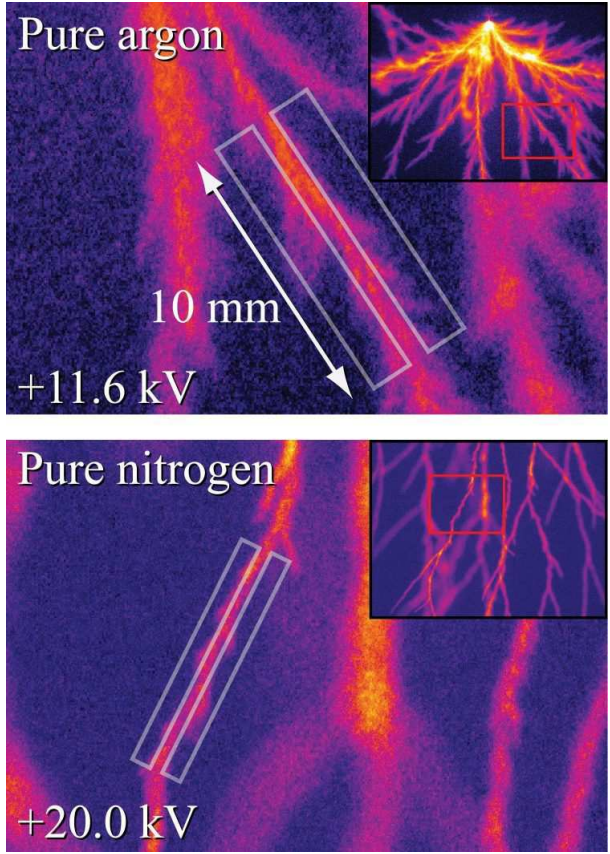}

\caption{\label{fig:Feather-structure}Feather like structures in a pure argon
and a pure nitrogen discharge. Both images are acquired with the C-supply
at 200~mbar. The images as captured by the camera are shown in the
boxes in the top right corners with the zoomed areas indicated. The
grey rectangles in the zoomed images indicate the area in which the
number of small branches was counted. The nitrogen image is on the
same scale as the argon image, but lower in the 160~mm gap (roughly
halfway).}

\end{figure}

We have analysed one channel from an argon image at 200~mbar (figure~\ref{fig:Feather-structure}).
We counted the number of side channels from a 10~mm long section
of this channel in a region between 0.5 and 2~mm from the centre
of the channel on both sides. Dividing this number by the volume of
the counting region (we assume cylindrical symmetry of the streamer
channel) leads to a branch density of about 10$^{2}$~cm$^{-3}$.

When we apply the same procedure to the image of a nitrogen discharge
from figure~\ref{fig:Feather-structure}, $ $we find roughly the
same branch density. Although the number of small side channels is
lower than in the argon image, so is their length. Therefore the counting
region reaches to only about 1~mm around the centre of the channel.
We will discuss the interpretation of this phenomenon in section~\ref{Feather-interpretation}.

\subsection{Time resolved images}

\begin{figure}
\includegraphics[width=1\columnwidth]{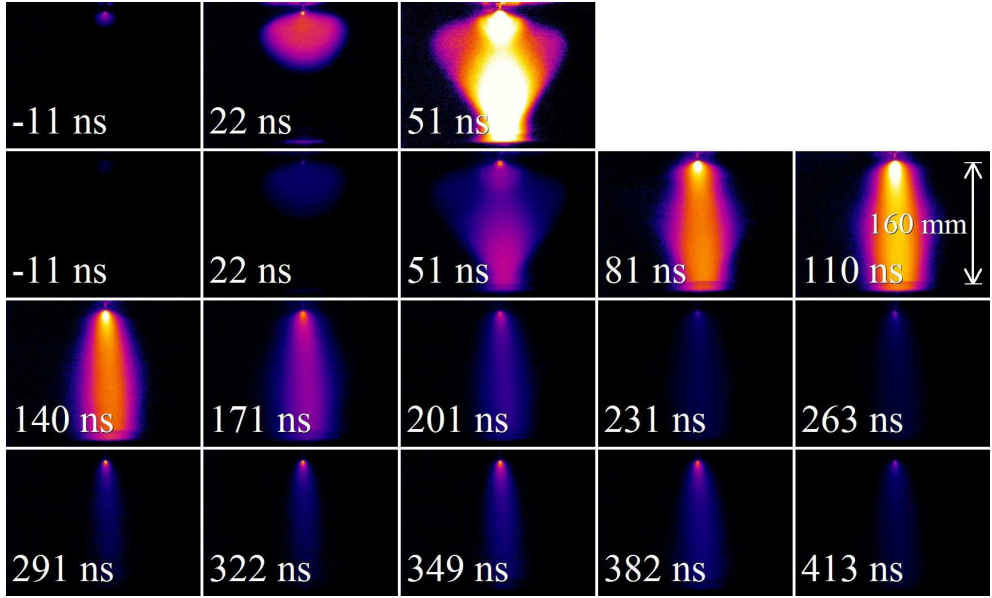}

\caption{\label{fig:Movie25mbar}Image sequence of streamer discharges produced
with a 20~kV pulse by the Blumlein pulser in pure nitrogen at 25~mbar.
The integration time of each image is 15~ns and the indicated time
is the camera delay with respect to the beginning of the voltage pulse.
The images in the top row are identical to the first three in the
second row, except for their colour representation. The images in
the top row have an \emph{Mf} value of 125, all other images have
an \emph{Mf} value of 8.6. An avi-movie of all recorded frames can
be found in the multimedia attachments.}

\end{figure}

\begin{figure}
\includegraphics[bb=0bp 0bp 1827bp 341bp,width=1\columnwidth]{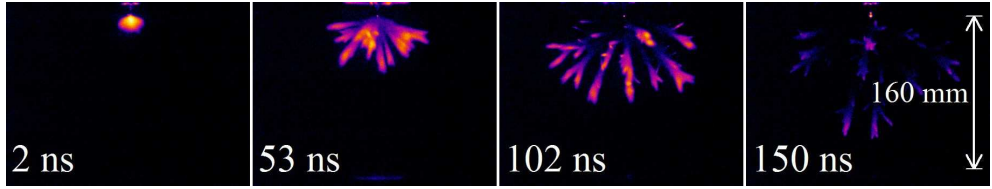}

\caption{\label{fig:Movie-100mbar}Image sequence of streamer discharges produced
with a 20~kV pulse by the Blumlein pulser in pure nitrogen at 100~mbar.
All images have an \emph{Mf} value of 158. Recall that we can not
capture more than one image per discharge event. Therefore the images
presented here are all from separate discharges. An avi-movie of all
recorded frames can be found in the multimedia attachments.}

\end{figure}

We took time resolved, movie-like image sequences by decreasing the
integration time of our camera and varying its internal delay value.
Results of these measurements are given in figures~\ref{fig:Movie25mbar}
and \ref{fig:Movie-100mbar}. All measurements were performed at about
20~kV with the Blumlein pulser as described in section~\ref{Blumlein-pulser}.
Note that our camera can only take one image per discharge event and
therefore these images are all from separate events. In figure~\ref{fig:Movie25mbar},
the streamer crosses the gap in less than 15~ns. After this time,
it evolves into a glow discharge, which is much brighter than the
original streamer and initiation cloud. This glow discharge remains
bright for the duration of the voltage pulse (130~ns), and then starts
to extinguish.

After roughly 290~ns, the discharge appears again. This is a negative
discharge caused by the reflected pulse in the Blumlein pulser (see
figure~\ref{fig:Comparison-pulse-shapes}). Obviously, at this pressure,
the gap is overvolted at 20~kV, which explains the high propagation
velocity of about $3\cdot10^{6}$~m/s.

At 100~mbar, the streamers do not reach the other side of the gap
within the pulse duration. Therefore, the light is only emitted by
the streamers and no glow discharge is formed, see figure~\ref{fig:Movie-100mbar}.
The images at 102 and 150~ns clearly show that the light is only
emitted by the propagating streamer heads, as was shown before by
Briels \emph{et al.} \cite{Briels2008b}.

\subsection{Effects of repetition frequency}

\begin{figure}
\includegraphics[width=9cm]{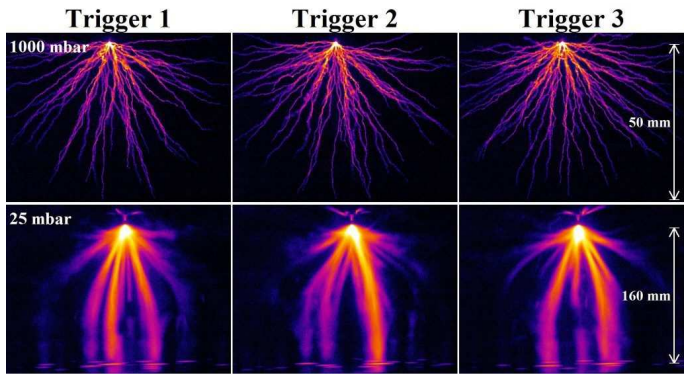}

\caption{\label{fig:Repetition-frequency}Discharge images as function of trigger
number. After a waiting period of about 30~s, the discharge pulse
generator is started with a repetition frequency of 1~Hz. The images
shown here are triggers 1, 2 and 3 after the waiting period. All images
are taken in pure nitrogen with the C-supply. The images in the top
row are zoomed images taken with 1000~mbar N$_{2}$, 40~kV and have
an \emph{Mf}-value of 350. The images in the bottom row are full images
taken with 25~mbar N$_{2}$, 15~kV and have an \emph{Mf}-value of
1.4. Note that the images in one row are not from the same series.
We have only stored one image per measurement series.}

\end{figure}
We have performed measurements on the effect of the pulse repetition
frequency with frequencies down to 0.03~Hz. An example of such a
measurement can be seen in figure~\ref{fig:Repetition-frequency}.
In this measurement, the power supply was triggered with a repetition
frequency of 1~Hz. After a measurement, the triggering was stopped,
and resumed 30~seconds later. Next, one image was stored, corresponding
to either the trigger 1, 2 or 3 after the waiting period. We have
not observed any influence of trigger number on streamer morphology
and other properties (all images in one row of figure~\ref{fig:Repetition-frequency}
are indistinguishable). This was tested and observed for pure nitrogen
and for the 0.2\% oxygen in nitrogen mixture. We have measured up
till trigger 5, but this leads to the same results as triggers 1,
2 and 3.

We have also seen that within a single series of subsequent voltage
pulses, streamers follow a new (semi-random) route on every new voltage
pulse. It seems that streamer channels do not have a tendency to follow
the same path twice nor is there any indication that they are in any
way influenced by the shape of the previous discharge (at 1~Hz repetition
frequency). This is of course not the case for conditions that lead
to only one streamer channel. In this case, the one channel always
is located on the symmetry axis of the experimental setup.

Unfortunately we do not have any images to prove this statement, as,
in our measurement procedure, images are stored manually and we have
not stored any image sequences. However, during a measurement session,
the images are shown on the screen with the repetition frequency of
1~Hz (this is then stopped to store an image). From our experience,
consecutive images on the screen are completely independent from each
other. Any tendency to follow old paths would be easily observable,
especially in situations with a limited number of streamer channels
(e.g. 3--50). In more than six years of streamer measurements by at
least seven different people, such behaviour has never been observed.\\

The only effect of repetition frequency that we have observed is,
that for certain conditions where streamer initiation is not easy
because the applied voltage is close to the threshold, the chances
of initiation are increased for repeated discharges. In these conditions,
the first image of a trigger series (trigger 1) is often completely
empty, or only contains a small glowing cloud around the electrode.
Subsequent triggers do then often lead to a complete streamer discharge.
It can be expected that the jitter (i.e. variations in initiation
time) of the first few discharges in these conditions is also large,
but we have not measured this. In most of our measuring conditions,
the initiation jitter was below 10~ns.

These effects regarding initiation may be attributed to leftover ionization
or metastables from previous discharges. As the discharge always starts
around the tip, any leftover ionization will have an influence on
the initiation and maybe the early propagation of the streamer. However,
we did not perform a detailed investigation of this phenomenon, as
we are focusing on bulk streamer propagation and not initiation or
other electrode effects.

\subsection{Minimal streamer width}

\begin{figure}
\includegraphics[width=0.9\columnwidth]{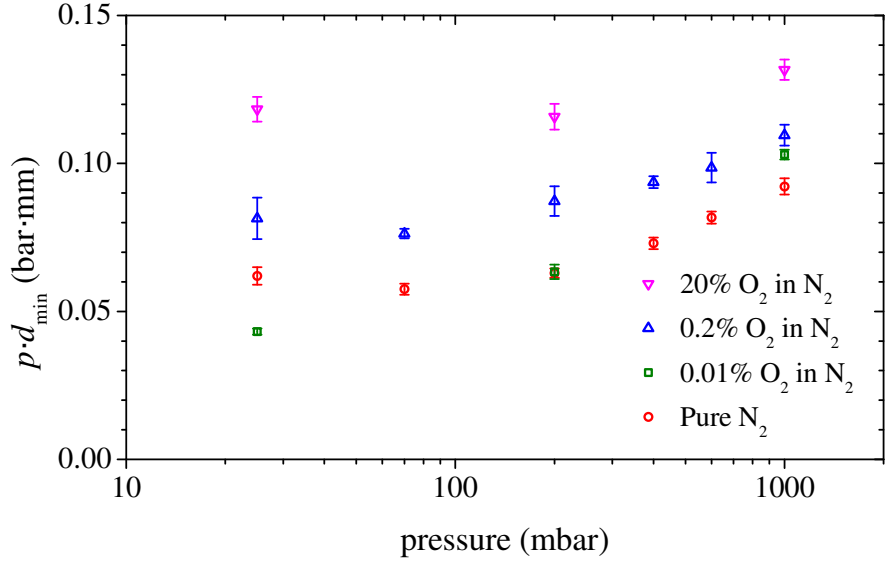}

\caption{\label{fig:pdminGraph}Scaling of the reduced diameter ($p\cdot d_{min}$)
with pressure ($p$) for the four different nitrogen oxygen mixtures.
Every point represents 4 to 10 measured streamer channels each from
a separate discharge image. The error bars give the standard error
and do not include systematic errors. All measurements are perfomed
at room temperature; Therefore we use pressure instead of density
to scale streamers.}

\end{figure}
Briels \emph{et al.} have also shown that the reduced streamer diameter
$n\cdot d_{min}$ is constant as function of the density $n$ for
a specific gas mixture \cite{Briels2008b}. The choice for these thinnest
channels has been made because experimental data as well as theoretical
considerations show that there is a lower limit in streamer diameter,
but no upper limit of the streamer diameter with growing voltage has
been found yet.

The minimal streamer diameters have been measured on images where
the streamer diameter (FWHM) is at least 10~pixels wide. We took
zoomed images like the 200~mbar images from figure~\ref{fig:Zoomed-images}
for this purpose, by moving the camera closer to the discharge, and
even higher magnifications at higher pressures. As stated before,
we were unable to reliably measure diameters for streamers in pure
oxygen and pure argon. Measurements of the minimal streamer diameters
are given in figure \ref{fig:pdminGraph}. All results have been obtained
with the C-supply. Some observations:
\begin{itemize}
\item The reduced diameter is roughly constant for any gas mixture, but
it does increase slightly at higher pressures.
\item The reduced diameter increases as a function of oxygen concentration
although there is no significant difference between pure nitrogen
and nitrogen with 0.01\% oxygen.
\end{itemize}
The increase of reduced diameter as function of pressure could be
a measurement artifact: it is possible that the width of the streamers
at high pressures is still overestimated because of bad focusing (shallow
depth of field) and other imaging artifacts \cite{Briels2006}.

Ono and Oda~\cite{Ono2003} also find an increase of diameter although
they report an increase of a factor 12 between pure nitrogen and a
20\% oxygen in nitrogen mixture. This large factor can be attributed
to the fact that Ono and Oda do not specifically look for minimal
streamers. They use a pulse of at least 15~kV in gas at atmospheric
pressure in a 13~mm point-plane gap. This is an overvolted gap, which
will lead to streamers that are thicker than the minimal streamers
we look at. We see as well that air streamers can be thicker than
pure nitrogen streamers, but our minimal streamers are similar for
all mixtures.

When we compare the values from figure~\ref{fig:pdminGraph} with
the results of Briels \emph{et al.~}\cite{Briels2008b} we see that
they are lower in all cases. Briels \emph{et al.} give an average
value of $p\cdot d_{min}$ of 0.12~bar$\cdot$mm for \textquoteleft{}pure\textquoteright{}
nitrogen and 0.20~bar$\cdot$mm for air. We have found 0.07~bar$\cdot$mm
and 0.12~bar$\cdot$mm respectively. There are several reasons for
the differences:
\begin{itemize}
\item Influences of other gas components like water vapour and carbon dioxide
that are more prevalent in the set-up of Briels than in the new high
purity set-up.
\item Small differences in voltage pulse rise time and amplitude that could
lead to larger than minimal streamers in the case of Briels.
\item Overestimation of streamer widths by Briels because of the optical
problems explained above. 
\end{itemize}
Note that the pure nitrogen in the work of Briels \emph{et al.} was
probably less pure than in the present work. The new set-up is much
better suited for work on high purity gasses than the old one used
by Briels \emph{et al.} It is estimated that the \textquoteleft{}pure\textquoteright{}
nitrogen of Briels \emph{et al.} has a contamination level of less
than 0.1\%, while we here achieve less than 0.0001\%.

\subsection{Velocity measurements}

\begin{figure}
\includegraphics[width=0.9\columnwidth]{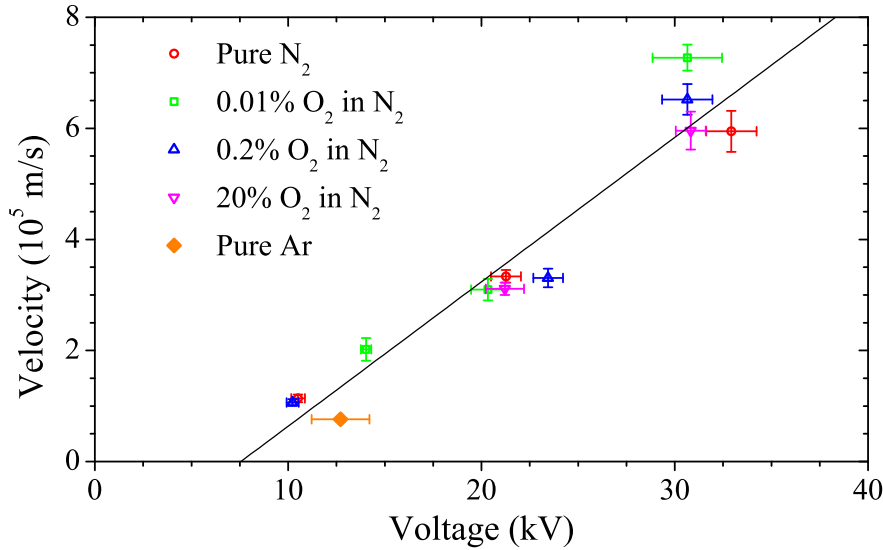}

\caption{\label{fig:Velocity-measurement-results}Velocity measurement results
at 200~mbar for five different gasses. The horizontal error bars
indicate the spread in the applied voltages, the vertical error bars
indicate the sample standard deviation for 4--6 measurements. The
velocities are measured roughly in the middle of the gap, except for
the cases where the discharge does not reach this far (at lower voltages),
in these cases, the velocity is determined at a position close to
the end of the streamer channels.}

\end{figure}

Streamer velocities have been determined on discharges in all nitrogen-oxygen
mixtures and in argon. In the nitrogen-oxygen mixtures, the length
of a streamer is determined with the first method discussed in section~\ref{Velocity-technique}
and demonstrated in figure~\ref{fig:Velocity-example}. For argon,
the second method must be used due to the long emission time after
impact excitation. Further, it must be noted that the usable voltage
range (for the C-supply) is limited to only a few kilo-volts when
using argon: if the voltage is too low there is no discharge at all
and if the voltage is too high we get a spark that can potentially
damage the camera. Therefore, we only have measured one point for
argon. All results have been obtained with the C-supply. The results
of the measurements at 200~mbar are given in figure~\ref{fig:Velocity-measurement-results}. 

The reason why we do not present measurements at 1000~mbar is that
we were not able to use voltages above 45~kV and therefore the streamers
extinguish after a short propagation time, as is shown in figure~\ref{fig:Overview-images}.
We also do not show measurements at 25~mbar. At this pressure, the
streamers are so wide, that no fully evolved streamer channel can
occur before the cathode plate is reached (see e.g. figure~\ref{fig:Movie25mbar}).

The most striking feature of these measurements is the fact that all
five gasses show very similar velocities and voltage dependence. This
is also observed at other pressures. The measured velocities have
a minimum value of $0.5-1\cdot10^{5}\,\mathrm{m/s}$. Although the
measurements at 1000~mbar and 25~mbar do not give a complete set
of results, like for 200~mbar, we can state that at these pressures
the same minimum velocity is found. We did not observe the increase
in propagation velocity with increasing oxygen fraction that is reported
by Yi and Williams~\cite{Yi2002}. The increase of velocity with
voltage is almost in quantitative accordance with the results found
by Briels~\cite{Briels2008}, although she published results measured
at 1000~mbar in a 40~mm gap, while we now focus on 200~mbar in
a 160~mm gap.

\section{Interpretation}

\begin{figure}
\includegraphics[width=1\columnwidth]{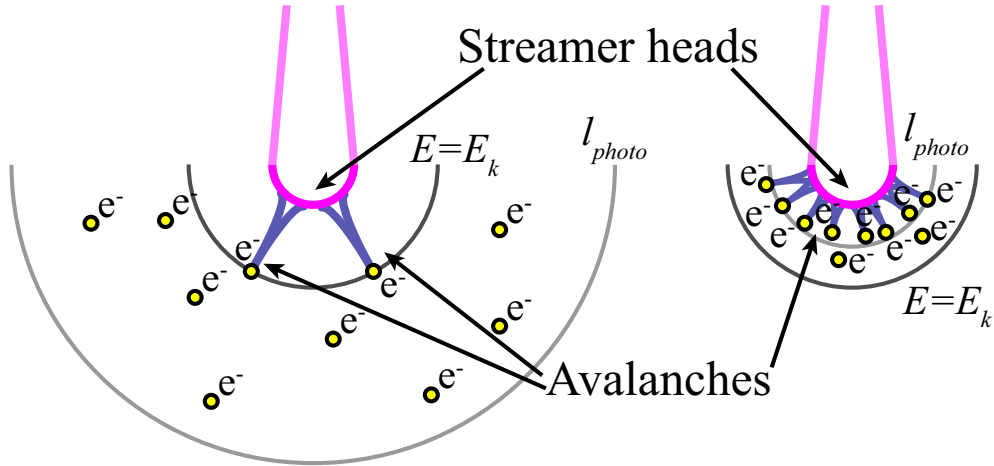}

\caption{\label{fig:Schematic-Feather}Sketch of the mechanism of feather formation.
On the left hand side, the photo-ionization length ($l_{photo}$)
is significantly larger than the active region where the local field
exceeds the breakdown field ($E>E_{k}$). The right hand side illustrates
the reversed situation.}

\end{figure}

The photo-ionization mechanism as described by Teich~\cite{Teich1967}
is applicable to oxygen-nitrogen mixtures: the nitrogen concentration
determines the number of photons in the 98 to 102.5~nm wavelength
range; within our experiments where the oxygen concentrations range
from below 1 ppm up to 20\%, the photon source can be considered as
essentially constant (if field enhancement and radius at the streamer
tip are the same).

The oxygen concentration determines the absorption length $l_{photo}$
of these photons, and therefore the range of availability of free
electrons. The absorption length is about 1.3~mm in air at standard
temperature and pressure, and it scales linearly with inverse oxygen
density; the oxygen density obviously depends both on the gas density
and on the N$_{2}$:O$_{2}$ mixing ratio~\cite{Luque2008b}.

\subsubsection*{Feather-like structures}

\label{Feather-interpretation}An interpretation for the feather-like
structure of the streamer channels in argon and in gas mixtures with
low oxygen concentrations is the following. The small branches that
form the feathers are in fact avalanches moving towards the streamer
head. This is illustrated in figure~\ref{fig:Schematic-Feather},
where two extreme cases are shown. In the first case, there is a low
O$_{2}$ concentration and therefore the photo-ionization length is
long. When this length is much larger than the region where the electric
field ($E$) exceeds the critical breakdown field $E_{k}$, ionization
is mainly created at places where the field is significantly below
$E_{k}$. Therefore the created free electrons can not form avalanches
immediately after creation, or they are lost in the lateral direction
where the field never exceeds $E_{k}$. Only when the streamer head
comes so close that $E\geq E_{k}$, avalanches form. These separate
avalanches are visible in the form of the feathers we see in the images
of argon and pure nitrogen.

On the other hand, if there is a sufficiently high oxygen concentration
in nitrogen to have a photo-ionization length that is significantly
below the radius where $E=E_{k}$, all electrons that are created
will immediately be accelerated towards the streamer head and create
avalanches. Because the density of the seed electrons is much higher
(the same amount of UV-radiation is absorbed in a much smaller volume),
these avalanches will overlap and become part of the propagating streamer
head. Therefore, the streamer will appear smooth, they easily will
become broad, and they will not form any feather-like structures.

If photo-ionization is not present at all, but if there is some background
ionization, the situation will be similar to the low photo-ionization
case. Now the electrons outside the $E\geq E_{k}$ region will not
be created by photo-ionization, but by other processes like cosmic
rays or radioactive decay (in air often radon escapes from building
materials). In our stainless steel vessel filled with pure gasses,
the amount of radioactive radon will be orders lower than in ambient
air.\\

Figure~\ref{fig:Schematic-Feather} largely resembles textbook concepts
of streamer branching due to avalanches caused by single electrons;
the first version of such plots stems from Raether~\cite{Raether1939}.
This concept is criticised by Ebert \emph{et al.} in section~5.3
and figure~10 of~\cite{Ebert2006a}. Here it is stated that the
concept that non-local photo-ionization leads to branching has never
been substantiated by further analysis and that even if this avalanche
distribution is realised, it has not been shown that it would evolve
into several new streamer branches. In contrast, it is stated that
the formation of a thin space charge layer is necessary for streamer
branching while stochastic fluctuations are not necessary, especially
because even in fully deterministic, but nonlinear models, instabilities
and branching are possible. These concepts are further elaborated
in~\cite{Montijn2006b,Luque2007,Derks2007,Tanveer2009,Kao2009}.

Concerning the emergence of feathers rather than branches: there are
stochastic avalanches, that we interpret as the \textquotedbl{}hairs
of feathers\textquotedbl{}, but the streamer is apparently not in
an unstable situation and does not branch. Of course, this statement
is partially semantic - one also could consider the hairs as little
side branches that immediately die out again. The real difference
between branches and avalanches lies in the question whether the hairs
do build up own space charges, or whether they just evolve in the
enhanced field of the main streamer. Figure~\ref{fig:Schematic-Feather}
suggests that the second is the case. They die out rather after the
electron avalanche has reached the main streamer and do not reach
a propagating state.\\

If this interpretation is correct, the number of small branches could
be a measure for the background ionization density when no photo-ionization
exists. The branch density in the measurements at 200~mbar in pure
nitrogen and argon is about 10$^{2}$~cm$^{-3}$. Unfortunately,
there is very little literature regarding background ionization in
pure gasses in containers. Estimations for the maximum equilibrium
charge density in air at atmospheric pressure are in the range of
$10^{3}-10^{4}\:\mathrm{cm^{-3}}$ \cite{Pancheshnyi2005}. However,
this value can mainly be attributed to ionization by $\alpha$ particles
emitted by decaying radon gas~\cite{Pancheshnyi2005}. When we assume
that the concentration of radon inside the closed metallic cylinder
of our experimental set-up is some orders below the ambient value
and that the pressure is 200~mbar instead of 1~bar, a background
charge concentration of 10$^{2}$~cm$^{-3}$ seems reasonable.

As was stated before, the first Townsend ionization coefficient of
argon is much higher than in molecular gasses (at low field strengths)
\cite{Siglo} because there are no rotational or vibrational states.
Therefore the region where $E\geq E_{k}$ is larger. This can explain
why the feather-like structures are more pronounced in argon discharges
than in nitrogen discharges.

\subsubsection*{Electron attachment by oxygen}

Besides photo-ionization, oxygen also plays another role since it
is an attaching gas, in contrast to nitrogen or argon. This means
that it removes electrons and therefore conductivity from the gas.
This mechanism comes in action both ahead of the streamer head and
in the streamer channel. Ahead of the streamer head and in the whole
non-ionized region, it binds free electrons and makes them essentially
immobile, but it also can release them again by detachment when the
field exceeds a critical value. In the streamer channel, it limits
the conductivity after a sufficiently long propagation time.

If, in our simple model of the feather structure (figure~\ref{fig:Schematic-Feather}),
the detachment field is higher than the critical breakdown field $E_{k}$,
then the electrons at the position $E=E_{k}$ can still be attached
to an oxygen molecule and the avalanches will start closer to the
streamer head. However, this will only occur at very low pressures.
The exact value of the detachment field depends on the vibrational
state of oxygen and will therefore be different for background ionization
than for photo-ionization. A detailed discussion of the role of detachment
can be found in~\cite{Pancheshnyi2005}.

Aleksandrov \emph{et al.}~\cite{Aleksandrov2001} also find that
the addition of 1\% of oxygen to a discharge in pure argon leads to
a faster decay of the streamer and requires a higher electrical energy
input. However, their simulations attribute this not to electron attachment
to oxygen, but to quenching of excited argon states by oxygen molecules.\\

A more quantitative analysis, including simulations, of the topics
discussed in this section is in preparation by Wormeester, Luque,
Pancheshnyi and Ebert.

\section{Conclusions}

The most simple and maybe surprising conclusion we can draw from the
experiments described here is that several important streamer properties
are quite similar for all nitrogen-oxygen mixtures and pure nitrogen
with an impurity level of less than 1~ppm. This applies to many of
the overview images and zoomed images, but more so for the minimal
streamer diameters and velocities. We have tested this on pure nitrogen
and three different nitrogen-oxygen mixtures with nearly six orders
of magnitude difference in oxygen fraction. We do see clear differences
when decreasing the oxygen concentration, but the streamers do still
propagate with roughly the same velocity and their minimal diameter
decreases by less than a factor of two from artificial air to pure
nitrogen.

This is remarkable because it means that either the photo-ionization
mechanism is not as important in streamer propagation as previously
thought, or that this mechanism can still provide enough free electrons,
even at very low oxygen concentrations. When we assume that the direct
photo-ionization mechanism is not the major source of free electrons
there are a few options left for these sources: 
\begin{itemize}
\item Other photo-ionization mechanisms than direct photo-ionization of
oxygen by nitrogen emission are responsible for free electrons. For
example, this could be step-wise ionization of nitrogen molecules,
though this is unlikely at the low electron densities and high propagation
velocities of the streamer tip. The measurements of Penney and Hummert~\cite{Penney1970}
in pure nitrogen and pure oxygen indicate that there is photo-ionization
in pure gasses, although in their case, the gasses are orders of magnitude
less pure than our gasses. 
\item Background ionization of the gas due to cosmic radiation, radioactivity
or leftover charges from previous discharges can deliver enough free
electrons for streamer propagation. The background ionization can
lead to free or bound electrons. The bound electrons can be detached
by the enhanced electric field of the streamer head \cite{Pancheshnyi2005}.
This can be tested by modifying the background charge density in some
way (e.g. adding radioactive isotopes) and studying its effect on
streamer properties. 
\end{itemize}
Numerical simulations of Pancheshnyi \cite{Pancheshnyi2005} show
that changing the background ionization of the gas by two orders of
magnitude from $10^{5}\:\mathrm{cm^{-3}}$ to $10^{7}\:\mathrm{cm^{-3}}$
results in only small changes (order 10--30\%) of streamer diameter,
current and propagation velocity. This is in line with both the hypothesis
that a very low amount of oxygen is enough to sustain photo-ionization
and the hypothesis that background ionization is the source of the
electrons needed for propagation. Note that the literature value for
the background ionization in ambient air is $10^{3}-10^{4}\:\mathrm{cm^{-3}}$.
From the density of feathers, we estimate $10^{2}\:\mathrm{cm^{-3}}$
in our 200~mbar pure gasses inside our metal vessel.

We have not observed any influence of repetition frequency (0.03 --
1~Hz) on streamer morphology and propagation. Neither have we observed
any tendency for streamers to follow the path of streamers from previous
discharges. Therefore we conclude that the influence of previous discharges
on the ionization density is either very small, or does not affect
streamer properties. \\

Yi and Williams~\cite{Yi2002} do find an oxygen-fraction dependent
streamer velocity. They claim that this is caused by the difference
in photo-ionization. An important difference between their measurements
and ours is that they have been performed at much higher voltages
(70~kV and above) and have a significantly different electrode geometry.
Like us, they find that the streamer diameter increases with oxygen
fraction. They again attribute this to photo-ionization. More photo-ionization
leads to more non-local electron production and thus to a wider streamer
head. Besides this, more photo-ionization would also reduce the required
field enhancement in the streamer head, which allows the streamer
head to have a blunter shape and reduces branching. Simulations by
Luque \emph{et al.} support this \cite{Luque2007}, although these
simulations treat negative streamers and we have looked at positive
streamers. More detailed descriptions and simulations of these mechanisms
are in preparation by Wormeester \emph{et al.}

Although there are some quantitative differences (mainly a $\sim$40\%
reduction in measured streamer diameter), the qualitative conclusions
of Briels \emph{et al.} \cite{Briels2008b} regarding the similarity
laws have been confirmed in this work, also for the gasses of much
higher garantueed purity investigated here.

\ack{}{S.N. acknowledges support by STW-project 06501, part of the Netherlands'
Organization for Scientific Research NWO. We like to thank Guus Pemen
and Zhen Liu for their help in the design of the Blumlein pulser and
Sergey Pancheshnyi and Gideon Wormeester for the discussions regarding
the interpretation of our experiments.}

\section*{References}

\bibliographystyle{unsrt}
\bibliography{puregasses}

\end{document}